\documentstyle[12pt,epsfig]{article}

\begin{document}

\begin{titlepage}
\begin{flushright}
UCLA/97/TEP/5\\
hep-ph/9702380 \\
February 19, 1997\\
\end{flushright}

\vskip 2.cm

\begin{center}
{\Large\bf Comment On Nonperturbative ${\cal O}(1/m_c^2)$ Corrections
to $\Gamma( \bar{B} \rightarrow X_s \gamma)$.}
\vskip 2.cm

{\large A. K. Grant, A. G. Morgan, S. Nussinov$^*$ and R. D. Peccei}

\vskip 0.5cm

{\it Department of Physics, University of California at Los
Angeles,\\Los Angeles, CA 90095-1547}
\vskip 1cm
{\it $^*$ School of Physics and Astronomy,\\ Raymond and Beverly Sackler 
Faculty of Exact Science,\\ Tel-Aviv University,\\ Tel-Aviv 69978, Israel}
\end{center}
\vskip 3cm

\begin{abstract}

{\baselineskip=24pt
We present an estimate of certain higher order corrections to the
contribution of the charm triangle loop in the inclusive
$\bar{B}\rightarrow X_s \gamma$ decay rate, recently discussed by
Voloshin.  We find that these corrections are minute and hence
the result found by Voloshin, although small, is quite robust.}

\end{abstract}
\vfill
\end{titlepage}

The concepts of heavy quark universality, symmetry and effective field
theory have been exhaustively applied to inclusive B
decays \cite{HQET}. By these means, the inclusive $\bar{B}\rightarrow
X_c\ell\nu$ and $\bar{B}\rightarrow X_s \gamma$ decays can be related to the
underlying (perturbatively computable) $b\rightarrow c \ell \nu$ and
$b\rightarrow s\gamma$ quark decays with only ${\cal O}(\Lambda_{\rm
QCD}^2/m_b^2)$ soft physics and $b$-quark binding corrections. These
results have been brought into question by a recent paper of Voloshin
\cite{Voloshin}, where a non-perturbative correction to the inclusive
rate for $\bar{B}\rightarrow X_s\gamma$ is identified which scales as
$\Lambda_{\rm QCD}^2/m_c^2$.

The appearance of non-perturbative corrections apparently missed in
the heavy quark effective theory (HQET) treatment of inclusive $B$
decays is unsettling and deserves a more thorough discussion. In this
note we show how, in the unphysical limit where $m_c^2 \gg
m_b\Lambda_{\rm QCD}$, it is possible to understand how the Voloshin
correction arises in the context of HQET. In this limit, in fact, this
correction is formally less than terms of ${\cal O}(m_c^2/m_b^2)$ and
hence is {\em under control}.  In real life, however, $m_c^2 \sim m_b
\Lambda_{\rm QCD}$ and, in principle, Voloshin's correction is subject
to considerable uncertainty.  In practice, however, we show that all
corrections to Voloshin's result are quite negligible, so that indeed
his computation provides the dominant contribution.

The corrections of ${\cal O}(\Lambda_{\rm QCD}^2/m_c^2)$ identified by
Voloshin \cite{Voloshin} arose by considering the contribution of the
gluon-photon penguin graph, shown in Fig. 1.  After applying a Fierz
transformation to the $\bar{s}_L \gamma_\mu c_L \bar{c}_L \gamma^\mu
b_L$ four-quark operator in the underlying effective Hamiltonian this
graph is proportional to the famous $AVV$ triangle diagram.  Because
of the GIM mechanism, there is no anomaly in the axial vertex, and the
result is necessarily non-singular in $(k_\gamma + k_g)^2$.  In the
limit of vanishing gluon momentum considered by Voloshin
\cite{Voloshin}, this contribution must scale as $V_{is}^* V_{ib} /
m_i^2$ for the up-type quarks $i$ running in the loop.  In his
calculation, Voloshin ignores the up quark loop because of its
miniscule CKM matrix elements and because, in no sense, can $m_u$ be
considered larger than the typical gluon momentum in the problem.
Since $V_{cs}^* V_{cb} \simeq -V_{ts}^* V_{tb}$ and $m_t^2 \gg m_c^2$,
the $c$ quark loop in Fig. 1 dominates and, indeed, the dominant
effective operator for the $b\rightarrow s \gamma g$ process scales as
$1/m_c^2$.  This is the origin of Voloshin's result for the effective
Lagrangian for this process:
\begin{equation}
\label{EffectiveLagrangian}
{\cal L}^{b\rightarrow s \gamma g}_{\rm Vol} = \frac{ e Q_c}{8 \pi^2}
\sqrt{2} G_F V_{cs}^* V_{cb} (\bar{s}_L \gamma^\mu \frac{\lambda^a}{2}
b_L ) \frac{ig}{3 m_c^2} G_a^{\nu \lambda} \partial_\lambda
\tilde{F}_{\mu\nu}.
\end{equation}
Other observations regarding effective Lagrangians of this type have
been made previously.  The effect on the exclusive $B\rightarrow K^*
\gamma$ decay has been studied in Ref.~\cite{Khodjamirian}, while
the authors of Ref.~\cite{Ali} have studied the effect of the
$b\rightarrow s\gamma g$ process on the photon energy spectrum.

\begin{figure}
\begin{center}
\epsfig{file=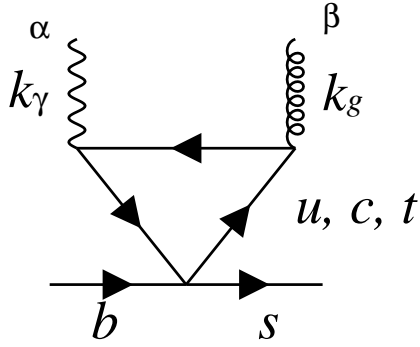,width=3in}
\caption{ Triangle diagram for the $b\rightarrow s \gamma g$ process.
We have omitted a second diagram with the gluon and photon
interchanged. }
\end{center}
\end{figure}

The interference of the above term with the leading perturbative
contribution for the $b\rightarrow s \gamma$ process, given by the
effective Lagrangian \cite{LO}
\begin{equation}
\label{LeadingOperator}
{\cal L}_{b\rightarrow s \gamma} = \frac{e}{16 \pi^2} \frac{4
G_F}{\sqrt{2}} V_{ts}^* V_{tb} C_7(\mu) m_b \bar{s}_L \sigma_{\mu\nu}
b_R F^{\mu\nu},
\end{equation}
in the operator product expansion (OPE) for the inclusive rate yields an
effective amplitude for the $b\rightarrow s\gamma$ process
\begin{eqnarray}
\label{EffectiveAmplitude}
T_{b\rightarrow s\gamma} &=&
- \frac{\alpha}{32\pi^4} G_F^2 m_b^5 \frac{1}{27} {\rm Re} \left[
	V_{cs}^* V_{cb} V_{ts} V_{tb}^* C_7
\right]  \left[
\frac{
	\bar{b} g \sigma_{\mu\nu} \frac{\lambda_a}{2}G_a^{\mu\nu} b
}{
	2 m_c^2
}
\right].
\end{eqnarray}
This produces a correction to the $\bar{B}\rightarrow X_s \gamma$ rate:
\begin{equation}
\label{VoloshinRate}
\frac{\delta \Gamma(\bar{B}\rightarrow X_s\gamma)}{\Gamma(\bar{B}\rightarrow
X_s\gamma)}
= \frac{1}{27 C_7} \frac{\mu_g^2}{m_c^2} \simeq -0.025.
\end{equation}
The numerical result above follows by using for the coefficient,
$C_7$, of the leading operator mediating the $\bar{B}\rightarrow
X_s\gamma$ transition, $C_7 \simeq -0.3$ \cite{CMM}, and using the
standard evaluation for the strength of the chromomagnetic interaction
of the $b$-quark inside the $B$ hadron \cite{BUV},
\begin{equation}
\mu_g^2 = \frac{1}{2} \langle B |
	\bar{b} g \sigma_{\tau\nu} G_a^{\tau\nu} \frac{\lambda_a}{2} b
|B \rangle = \frac{3}{4} ( {M_B^*}^2 -M_B^2 ) \simeq 0.4~{\rm GeV}^2.
\end{equation}

Although (\ref{VoloshinRate}) is a small correction to the inclusive
rate, its sensitivity to the scale $m_c^2$ brings into question the
expectations from HQET. To understand what is going on, it is useful
to examine more fully the gluon-photon penguin graph considered by
Voloshin. Because the AVV graph has been analyzed in detail by Adler
\cite{Adler}, it is straightforward to consider corrections to the
effective Lagrangian (\ref{EffectiveLagrangian}) or, equivalently, the
amplitude (\ref{EffectiveAmplitude}).

The triangle graph in question is proportional to the tensor
\begin{equation}
I_{\mu \alpha\beta}
=
I(k_\gamma, k_g) \left\{
	\epsilon_{\mu \nu \alpha \beta} (k_g\cdot k_\gamma)
	(k_\gamma^\nu - k_g^\nu)
	- \left[
		\epsilon_{\nu\mu\alpha\tau} k_{\gamma\,\beta}
		-\epsilon_{\nu\mu\beta\tau} k_{g\,\alpha}
	\right] k_\gamma^\tau k_g^\nu
\right\},
\end{equation}
where the invariant function, $I(k_\gamma, k_g)$, is given by
\begin{equation}
\label{InvariantFunction}
I(k_\gamma, k_g) = 24 \int_0^1 x\,dx \int_0^{1-x} y\,dy
\frac{1}{\left[
	m_i^2 - k_g^2 x(1-x) - 2xy k_g\cdot k_\gamma
\right]},
\end{equation}
with $m_i$ being the mass of the quark in the loop. In the limit where
the gluon $4$-momentum vanishes, Eq.~(\ref{InvariantFunction}) reduces
to the $1/m_c^2$ factor alluded to earlier. In fact, the gluon
$4$-momentum is never vanishing. It is of order of the typical
momentum of the $B$ meson constituents, which is of ${\cal
O}(\Lambda_{\rm QCD})$.\footnote{To be more precise, as we will argue
later, $|{\bf k}_g| \sim {\cal O}(\Lambda_{\rm QCD})$.}
Obviously, for the $u$-quark loop, it makes no sense to consider the
$k_g^\mu \rightarrow 0$ limit and so the scale $1/m_u^2$ never enters
the problem. Given the very small magnitude of $V_{us}^* V_{ub}$
relative to $V_{cs}^* V_{cb}$, it is perfectly sensible---as Voloshin
\cite{Voloshin} does---to neglect the $u$-quark loop altogether.

The situation is also clear for the $t$-quark loop. In this case $m_t$
is really much larger than all other scales. Given that the
$k_g^\mu \rightarrow 0$ limit is appropriate, the photon-gluon
penguin graph involving the $t$-loop will contribute an effective
interaction like that of Eq.~(\ref{EffectiveLagrangian}), but
scaled by $1/m_t^2$.  This interaction, however, gives a negligibly
small correction to the $\bar{B}\rightarrow X_s \gamma$ rate.  For the
$c$-quark loop, however, the effective $k_g^\mu \rightarrow 0$ limit
which yields the $b\rightarrow s\gamma g$ Lagrangian of
Eq.~(\ref{EffectiveLagrangian}) does not appear so safe.

It is certainly possible to drop the $k_g^2 x(1-x)$ term in the
denominator of Eq.~(\ref{InvariantFunction}) relative to $m_c^2$, since
$m_c^2 \gg \Lambda_{\rm QCD}^2$.  However, in the $B$ rest frame 
$|{\bf k}_\gamma| \sim m_b/2$, thus the $2 xy k_g\cdot k_\gamma$ term in
Eq.~(\ref{InvariantFunction}) is of ${\cal O}(m_b \Lambda_{\rm QCD})$,
which is not really small compared to $m_c^2$.  Voloshin's result
(Eq.~(\ref{EffectiveLagrangian})) neglects this term altogether.

Before evaluating corrections to Voloshin's formula, it is useful to
make some general remarks.  The approximation of also dropping the $2
xy k_g\cdot k_\gamma$ terms in Eq.~(\ref{InvariantFunction}) for the
$c$-quark loop is only tenable in a world where $m_c^2 \gg
\Lambda_{\rm QCD} m_b$. If this were the case, the Voloshin
correction to the $\bar{B}\rightarrow X_s \gamma$ inclusive rate, given in
Eq.~(\ref{VoloshinRate}), would not violate, per se, the heavy quark
expansion, since
\begin{equation}
\frac{\delta \Gamma(\bar{B}\rightarrow X_s\gamma) }{\Gamma(\bar{B}\rightarrow
X_s\gamma)} \sim \frac{\mu_g^2}{m_c^2} \sim \frac{\Lambda_{\rm
QCD}^2}{m_c^2} \ll \frac{m_c^2}{m_b^2}.
\end{equation}
On the other hand, dropping the $2 xy k_g\cdot k_\gamma$ term is {\bf
never} a good approximation in the heavy $m_b$ limit ($m_b \rightarrow
\infty$, with $m_c$ fixed) envisaged in the HQET.  In this limit, the
effective Lagrangian (\ref{EffectiveLagrangian}) scaled by $1/m_c^2$
never arises, since the proper limit for Eq.~(\ref{InvariantFunction})
is not $1/m_c^2$ but $-6/k_g\cdot k_\gamma$.  Indeed, in this limit the
photon-gluon penguin contribution vanishes identically because of the
GIM mechanism, $ V^*_{is} V_{ib} = 0$.

A class of systematic corrections to Voloshin's result can be computed
by retaining the full function $I(k_\gamma,k_g)$ in the photon-gluon
penguin graph. To be more precise, since $m_c^2 \gg \Lambda_{\rm
QCD}^2$, it suffices to consider
\begin{equation}
I(k_\gamma, k_g) = 24 \int_0^1 x \, dx \int_0^{1-x} y \, dy
\frac{1}{\left[
	m_c^2 - 2 xy k_\gamma\cdot k_g
\right]}.
\end{equation}
Expanding the denominator in powers of $k_\gamma\cdot
k_g/m_c^2$ identifies a progressive set of higher dimensional
operators which contribute to the $b\rightarrow s\gamma g$ Lagrangian.
This expansion effectively replaces the $G_a^{\nu\lambda}
\partial_\lambda \tilde{F}_{\mu \nu}/m_c^2$ operator in
Eq.~(\ref{EffectiveLagrangian}) by
\begin{eqnarray}
\frac{1}{m_c^2} G_a^{\nu \lambda} i\partial_\lambda
\tilde{F}_{\mu\nu}
&\rightarrow&
\frac{1}{m_c^2}
G_a^{\nu\lambda} i\partial_\lambda \tilde{F}_{\mu\nu}\nonumber\\
&+&\frac{4}{15 m_c^4} (i\partial_\alpha i\partial_\lambda
\tilde{F}_{\mu\nu})( iD^\alpha G_a^{\nu\lambda})\nonumber\\
&+& \frac{3}{35 m_c^6} (i\partial_\alpha i\partial_\beta i\partial_\lambda
\tilde{F}_{\mu\nu}) (iD^\alpha iD^\beta G_a^{\nu\lambda})
+ \ldots
\end{eqnarray}
The interference of these higher dimensional terms with the leading
order operator (\ref{LeadingOperator}) for the $b\rightarrow
s\gamma$ process in the operator product expansion formula for the
inclusive rate provides the desired set of corrections to the Voloshin
result.

\begin{figure}
\begin{center}
\epsfig{file=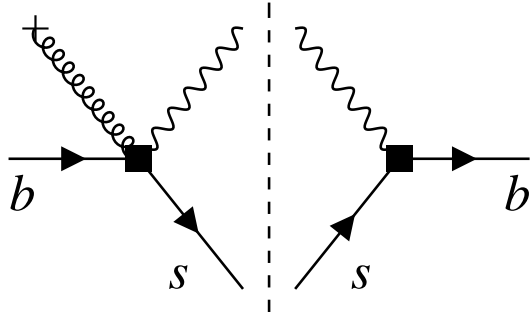,width=3in}
\caption{ The cut diagram which, together with its complex conjugate,
yields the contribution of ${\cal L}_{b\rightarrow s \gamma g}$ to the
inclusive $\bar{B} \rightarrow X_s \gamma$ rate. }
\end{center}
\end{figure}

In practice, it is simpler to compute these corrections by first
calculating the correlator of the leading order contribution to
$b\rightarrow s\gamma$, Eq.~(\ref{LeadingOperator}), with the full
$c$-quark photon-gluon penguin graph and then expanding the result in
powers of the gluon $4$-momentum.  A tedious but straightforward
calculation of the correlator depicted in Fig. 2 yields the
following formula for the effective amplitude for the $b\rightarrow
s\gamma$ process:
\begin{equation}
\label{full_result}
T_{b\rightarrow s\gamma} = - \frac{\alpha}{32 \pi^4} G_F^2 m_b^5
\frac{1}{27} {\rm Re} \left[
	V_{cs}^* V_{cb} V_{ts} V_{tb}^* C_7
\right] \; \bar{b} g \sigma_\nu^{\;\alpha} G_a^{\nu\lambda}
\frac{\lambda_a}{2} b J_{\alpha\lambda}.
\end{equation}
Here $J_{\alpha \lambda}$ involves an integral of $I(k_\gamma,k_g)$
over the photon and $s$-quark phase space and is given by
\begin{equation}
J_{\alpha\lambda} = -\frac{48\pi}{m_b^2}
\int \frac{d^3s}{(2\pi)^3\,2s^0}
\frac{d^3k_\gamma}{(2\pi)^3\,2k_\gamma^0}
(2\pi)^4 \delta^4(P-s-k_\gamma)
k_{\gamma\,\alpha}
k_{\gamma\,\lambda}
I(k_\gamma, k_g),
\end{equation}
where $P$ is the $b$ quark 4-momentum.

The tensor $J_{\alpha\lambda}$ can be expanded in terms of scalar functions 
which depend on the invariants which are left over after the phase
space integrations ($k_g^2$, $P\cdot k_g$, and $P^2$):
\begin{equation}
\label{decomposition}
J_{\alpha\lambda} = J_1 \eta_{\alpha\lambda} + J_2 P_\alpha P_\lambda
	+ J_3 k_{g \alpha} k_{g \lambda } + J_4 ( k_{g\alpha} P_\lambda +
	k_{g\lambda} P_\alpha).
\end{equation}
The scalar functions $J_i$ are easily identified by contracting 
$J_{\alpha\lambda}$ with the gluon and/or the $b$ quark 4-momentum.
They involve combinations of the phase space integrals of
$I(k_\gamma, k_g)$ , $(k_\gamma \cdot k_g ) I(k_\gamma, k_g)$ , and
 $(k_\gamma \cdot k_g )^2 I(k_\gamma, k_g)$.  These integrals are readily
computed in a power series in the gluon momentum. We will illustrate this 
with the phase space integral of $I(k_\gamma, k_g)$, which we perform
in the $b$ rest frame:
\begin{eqnarray}
\langle I(k_\gamma,k_g) \rangle &=& \int \frac{ d^3 s }{(2\pi)^3 2 s^0 } 
\frac{ d^3 k_\gamma }{(2 \pi)^3 2 k_\gamma^0 } (2\pi)^4 \delta^4(P-s-k_\gamma)
\nonumber\\
&\times& 24 \int_0^1 x\,dx \int_0^{1-x} y\,dy 
\frac{ 1 }{m_c^2 - 2 x y k_g \cdot k_\gamma }\nonumber\\
&=& -\frac{3}{2 \pi} \int_0^1 x\,dx \int_0^{1-x} y\,dy 
\frac{1}{ x y m_b |{\bf k_g}| }\ln
\frac{ 1 - \frac{ m_b ( E_g + |{\bf k_g}| ) x y }{m_c^2}}
     { 1 - \frac{ m_b ( E_g - |{\bf k_g}| ) x y }{m_c^2}}.
\end{eqnarray}
The gluon energy in the above is typically much smaller than the
gluon momentum.  Roughly speaking, the $B$ matrix elements involving
the gluon energy are of order the difference in kinetic energy of the
$b$ quark before and after absorbing a soft gluon.  Hence the gluon
energy is of order $E_g \sim \Delta T_b \sim \Lambda_{\rm QCD}^2 /
m_b$, while $|{\bf k}_g| \sim \Lambda_{\rm QCD}$--the gluons are
predominantly spacelike.  Dropping the gluon energy $E_g$ relative to
$|{\bf k}_g|$ and expanding the logarithm in powers of $m_b
\Lambda_{\rm QCD} / m_c^2$ yields a rapidly converging power series
for $\langle I(k_\gamma, k_g)\rangle$:
\begin{equation}
\label{expansion}
\langle I(k_\gamma, k_g)\rangle = \frac{1}{8\pi m_c^2}
\biggl{[} 1 + \frac{1}{140} \frac{m_b^2 |{\bf k}_g|^2 }{m_c^4} 
+ \frac{1}{6930}\frac{m_b^4  |{\bf k}_g|^4 }{m_c^8} + \ldots \biggr{]}.
\end{equation}
The higher order terms in the series involve integrals of $x^n y^n$,
which fall off asymptotically as $\sqrt{ \pi } / n^{3/2} 4^{n+1}$.

Using the decomposition (\ref{decomposition}) in Eq.~(\ref{full_result})
and retaining only the terms that do not vanish by the equations of motion 
gives for the $b\rightarrow s \gamma$ amplitude the expansion
\begin{eqnarray}
\label{amplitude}
T_{b \rightarrow s \gamma} & = & - \frac{\alpha}{32\pi^4} G_F^2 m_b^5
\frac{1}{27} {\rm Re} \left[ V_{cs}^* V_{cb} V_{ts} V_{tb}^* C_7
\right] \nonumber\\ &\times & \bar{b} \biggl{[} J_1\,
\sigma_{\nu\lambda}  + J_3\, 
\sigma_{\nu\alpha} k_g^\alpha k_{g\lambda}  + J_4\,
\sigma_{\nu\alpha} k_g^\alpha P_\lambda \biggr{]} g G_a^{\nu\lambda}
\frac{\lambda_a}{2} b
\end{eqnarray}
Calculations analogous to those that led to Eq.~(\ref{expansion}) lead
to the following expressions for the leading behavior of the scalar
functions $J_1$, $J_3$, and $J_4$ in the $b$ rest frame \footnote{The
$J_2$ term drops out because of the equations of motion.}:
\begin{eqnarray}
\label{scalar_coefficients}
J_1 &=& \frac{1}{2 m_c^2} \biggl{[} 1 + \frac{3}{700} 
				\frac{m_b^2 |{\bf k}_g|^2}{m_c^4} \biggr{]}
	\nonumber\\
J_3 &=& \frac{1}{2 m_c^2} \biggl{[} -\frac{3}{350} 
	\frac{m_b^2}{m_c^4} \biggr{]}
	\nonumber\\
J_4 &=& \frac{1}{2 m_c^2} \biggl{[} \frac{2}{15} \frac{1}{m_c^2} \biggr{]}.
\end{eqnarray}
Although the $J_4$ term in Eq.~(\ref{amplitude}) is nominally linear
in the gluon 4-momentum, it actually gives a correction of ${\cal O}
(\Lambda_{\rm QCD}^2/m_c^2)$ upon using Faraday's law for the gluon
fields (${\bf k \times E_a }= E_g {\bf B_a}$), since $E_g
\sim\Lambda_{\rm QCD}^2/m_b$.  Thus the leading correction to
Voloshin's result for $T_{b\rightarrow s \gamma}$ given in Eq.~(3)
arises only from the $J_1$ and $J_3$ terms and involves operators
quadratic in the gluon 4-momentum.  Using
Eq.~(\ref{scalar_coefficients}), one identifies this correction as
\begin{eqnarray}
\label{LeadingCorrection}
\delta T_{b \rightarrow s\gamma}& = &
- \frac{\alpha}{32\pi^4} G_F^2 m_b^5 \frac{1}{27} {\rm Re} \left[
V_{cs}^* V_{cb} V_{ts} V_{tb}^* C_7 \right]\nonumber\\
&\times& \left[ -\frac{3 m_b^2}{1400 m_c^6} \langle B| g \bar{b} 
\frac{\lambda_a}{2} ( \sigma_{\nu\lambda} (iD)^2 G_a^{\nu\lambda} + 
\sigma^{\nu\alpha} \{iD_\lambda , iD_\alpha \} G_a^{\nu\lambda} ) 
b | B \rangle\right].
\end{eqnarray}
Although one cannot relate the above matrix element to a property of
the $B$ mesons, as was done for the leading matrix element, we expect
it to be of ${\cal O} (\Lambda_{\rm QCD}^4)$.  However, because of the
tiny numerical coefficient, even though $m_b^2 \Lambda_{\rm QCD}^2\sim
m_c^4$, the above correction is totally negligible.  So Voloshin's
result~(\ref{VoloshinRate}) is robust, at least as far as these
corrections go.

There are, of course, many other corrections to Voloshin's result.
However, these are all of ${\cal O} (\Lambda_{\rm QCD}^4 / m_c^4)$,
down by a factor of $\Lambda_{\rm QCD}^2 / m_c^2$ relative to the
leading term calculated by Voloshin.  Some of these arise by
considering the full photon-gluon penguin graph, e.g. the $J_4$ term
in Eq.~(\ref{LeadingCorrection}).  Others come from terms where ${\cal
L}^{b\rightarrow s \gamma g}_{\rm Vol}$ is correlated with itself in
the OPE for the total width.  Yet others come from photon-gluon
penguin graphs in which two or more soft gluons are emitted.  Although
all of these contributions involve unknown matrix elements, because
$\Lambda_{\rm QCD}^2 \ll m_c^2$ it is reasonable to expect that they
also should yield quite small corrections to Eq.~(\ref{VoloshinRate}).

Our analysis gives some confidence that the $1/m_c^2$ nonperturbative
corrections to the $\bar{B} \rightarrow X_s \gamma$ process calculated
by Voloshin \cite{Voloshin} are an accurate estimate of these
corrections.  Unfortunately, these interesting corrections are
numerically small for this particular decay.  Their existence,
however, raises the interesting question of whether other $1/m_c^2$
corrections may give more sizable contributions to other processes.
We hope to return to this issue at a later date.

\begin{center} 
{\bf Acknowledgments}
\end{center}

We are grateful to Mark Wise for useful discussions on these matters.
He and his collaborators have arrived at similar conclusions to ours.
[See Z. Ligeti, L. Randall, and M. Wise, CALT-68-2097, MIT-CTP-2611,
hep-ph/9702322.]  We thank them for discussing their results prior to
publication.  We also would like to thank Misha Voloshin for a helpful
discussion.  One of us (S.N.) is grateful for the hospitality of the
UCLA department of Physics and Astronomy, where part of this work was
completed.  This work was supported in part by Department of Energy
grant DOE-FG03-91ER40662, Task C.

\end{document}